\documentclass[aip,apl, showpacs, reprint, preprintnumbers,amsmath,amssymb,superscriptaddress,floatfix]{revtex4-1}

\usepackage[]{graphicx}      
\usepackage{bm}             	
\usepackage{times}			
\usepackage{tabularx}
\usepackage{color}
\usepackage{dcolumn}		
\usepackage{amsmath}
\usepackage{amssymb}
\usepackage{amsfonts}
\usepackage{times}
\usepackage{tabularx}
\usepackage{xcolor,colortbl}

\begin{document}
\title{Ferromagnetic resonance study of composite Co/Ni - FeCoB free layers with perpendicular anisotropy}
\author{T. Devolder}
\email{thibaut.devolder@u-psud.fr}
\affiliation{Centre for Nanoscience and Nanotechnology, CNRS, Univ. Paris-Sud, Universit\'e Paris-Saclay, 91405 Orsay, France}
\author{E. Liu}
\affiliation{imec, Kapeldreef 75, 3001 Heverlee, Belgium}
\affiliation{Department of Electrical Engineering (ESAT), KU Leuven, Leuven 3001, Belgium}
\author{J. Swerts}
\author{S. Couet}
\author{T. Lin}
\author{S. Mertens}
\author{A. Furnemont}
\author{G. Kar}
\affiliation{imec, Kapeldreef 75, 3001 Heverlee, Belgium}
\author{J. De Boeck}
\affiliation{imec, Kapeldreef 75, 3001 Heverlee, Belgium}
\affiliation{Department of Electrical Engineering (ESAT), KU Leuven, Leuven 3001, Belgium}

\date{\today}                                           
%
%
\begin{abstract}
We study the properties of composite free layers with perpendicular anisotropy. The free layers are made of a soft FeCoB layer ferromagnetically coupled by a variable spacer (Ta, W, Mo) to a very anisotropic [Co/Ni] multilayer embodied in a magnetic tunnel junction meant for spin torque memory applications. For this we use broadband ferromagnetic resonance to follow the field dependence of the acoustical and optical excitation of the composite free layer in both in-plane and out-of-plane applied fields. The modeling provides the interlayer exchange coupling, the anisotropies and the damping factors. The popular Ta spacer are outperformed by W and even more by Mo, which combines the strongest interlayer exchange coupling without sacrificing anisotropies, damping factors and transport properties.
\end{abstract}

\maketitle

%
%

Magnetic tunnel junctions (MTJ) based on perpendicular magnetic anisotropy (PMA) systems are under active development for the next generations of spin transfer torque (STT) magnetic random access memories (MRAM). In this technology, the information is stored in the magnetization state of a free layer composed typically of Fe-rich FeCoB alloys sandwiched between Ta and MgO (so-called ''single'' MgO free layer\cite{ikeda_perpendicular-anisotropy_2010}) or between two MgO layers (''dual'' MgO free layers \cite{sato_perpendicular-anisotropy_2012}). Each MgO interface provides an interface energy that promotes PMA. The dual MgO option is gradually becoming more frequent for memory applications as it provides up to double anisotropy, hence more resilience to thermal fluctuations. This strategy ensures scalability \cite{sato_properties_2014, kim_ultrathin_2015} for junctions down to 20-30 nm of diameter but material solutions have to be found to scale further. To match with STT-MRAM objectives, one possible configuration  comprises (i) an Fe-rich, bcc-structured, low damping FeCoB at the MgO interface (ii) coupled ferromagnetically with a material supplying a strong anisotropy energy while maintaining damping and thickness as low as possible.

Among the material systems providing large PMA, the Co/Ni multilayers belong to the few ones that have a reasonably low damping, from  \cite{chen_spin-torque_2008} 0.033 down to \cite{song_relationship_2013, song_observation_2013} 0.021 and even\cite{haertinger_properties_2013} 0.014 depending on compositions and thicknesses. The unresolved questions are whether these low damping values can be maintained in ultrathin Co/Ni multilayers and whether the fcc-based Co/Ni multilayer can be coupled strongly to the bcc-based FeCoB layers across a texture transition.

In this paper, we compare three different spacers and evaluate their impact on the anisotropies, the interlayer exchange coupling and the damping in each layer of an MgO/FeCoB/spacer/[Co/Ni]$_{\times 4}$ /Pt system embodied in an MTJ. The studied spacers include the popular\cite{cuchet_influence_2014} Tantalum, as well as Molybdenum and Tungsten spacers. The choice of bcc metals (Ta, W and Mo) is to promote the fcc to bcc texture transition. The refractory character of W, Ta and Mo is also a foreseen advantage since large resilience to atomic diffusion upon annealing is desirable for CMOS back-end-of-line compatibility. W/CoFeB and Mo/CoFeB systems have indeed proven large resistance to annealing \cite{liu_thermally_2014, liu_high_2016} which correlate with the very slow diffusion of refractory metals in Fe \cite{powers_diffusion_1959, takemoto_diffusion_2007, nitta_diffusion_2002}. To assess the performance of these spacers, we use broadband ferromagnetic resonance (FMR) to follow the field dependence of the acoustical and the optical excitations of the composite free layer. The modeling provides the interlayer exchange coupling, the anisotropies and the damping factors. The popular Ta spacer is outperformed by Mo, which combines the strongest interlayer exchange coupling with no detrimental effect on the anisotropies and the damping within the magnetic materials of the stack. This qualifies Molybdenum as a material of choice for the spacer layer in composite free layers.

Our objective is to study hybrid free layers. By ''hybrid'' we mean comprising a bcc FeCoB layer which ensures optimal transport properties, coupled ferromagnetically to an fcc [Co/Ni] multilayer whose anisotropy strengthens thermal stability of the whole free layer. The multilayer is grown first on a Pt buffer that provides low coercivity and high anisotropy~\cite{liu_[co/ni]-cofeb_2016}. We then grow top-pinned MTJs of the following configuration \cite{liu_[co/ni]-cofeb_2016}: Pt  / [Co{(3\r{A})/}Ni{(6\r{A})]}$_{\times 4}$ ($t_2=3.5$ nm)  / spacer  / Fe$_{60}$Co$_{20}$B$_{20}$ ($t_1=1$~nm) / MgO / reference layer / cap. The [Co/Ni] multilayer is terminated by a nickel layer in contact with the spacer. The studied spacers are Ta(3\r{A}), Mo(3\r{A}) and W(3\r{A}). In addition, thicker (i.e. 5\r{A}) Mo and W spacers were used to decouple the two parts of the free layer and thereby measure of their respective moments and easy axes. W(5\r{A}) leads to in-plane magnetization of the FeCoB layer (not shown). We aim to optimize the spacer within hybrid free layers in \textit{realistic} MTJÕs, i.e. comprising reference layers that might influence the crystallization within the free layer. Our reference system \cite{liu_[co/ni]-cofeb_2016} is a standard Co/Pt based synthetic ferrimagnet (SAF). All samples were annealed at 300$^{\circ}$C for 30 min. in a field of 1 T.

%
\begin{figure*}
\includegraphics[width=14 cm]{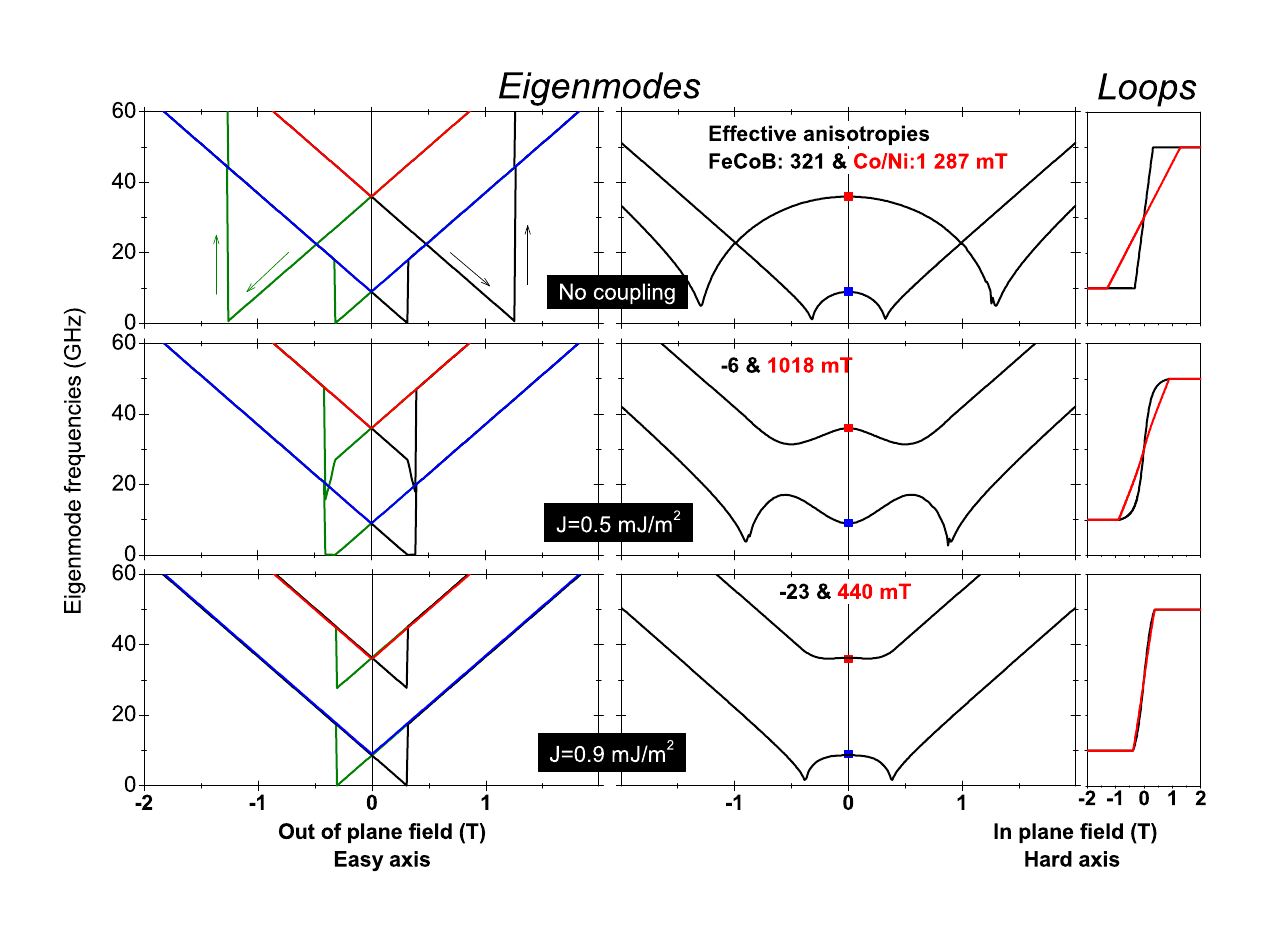}
\caption{(Color online). Eigenmode frequencies and hard axis loops for 3 hybrid free layers that share the same calculated eigenexcitation frequencies at remanence. The anisotropies and the interlayer couplings are chosen to get the eigenmode frequencies of 9 GHz (blue dot) and 36 GHz (red dot) at remanence like in the case of Mo(3\r{A}) spacer. Left column: eigenmode frequencies versus out-of-plane fields for an increasing (black) or decreasing (green) field sweep. In practice the layers' magnetizations switch near zero field and thus visit mainly the colored branches that correspond to the most stable states. Middle column: \textit{idem} versus in-plane fields. Right column: normalized hard axis loops (i.e. $M_x$ versus $H_x$) of the FeCoB layer (black lines) and the [Co/Ni] multilayer (red lines).}
\label{figure_methods}
\end{figure*}

\begin{figure}
\includegraphics[width=8.7 cm]{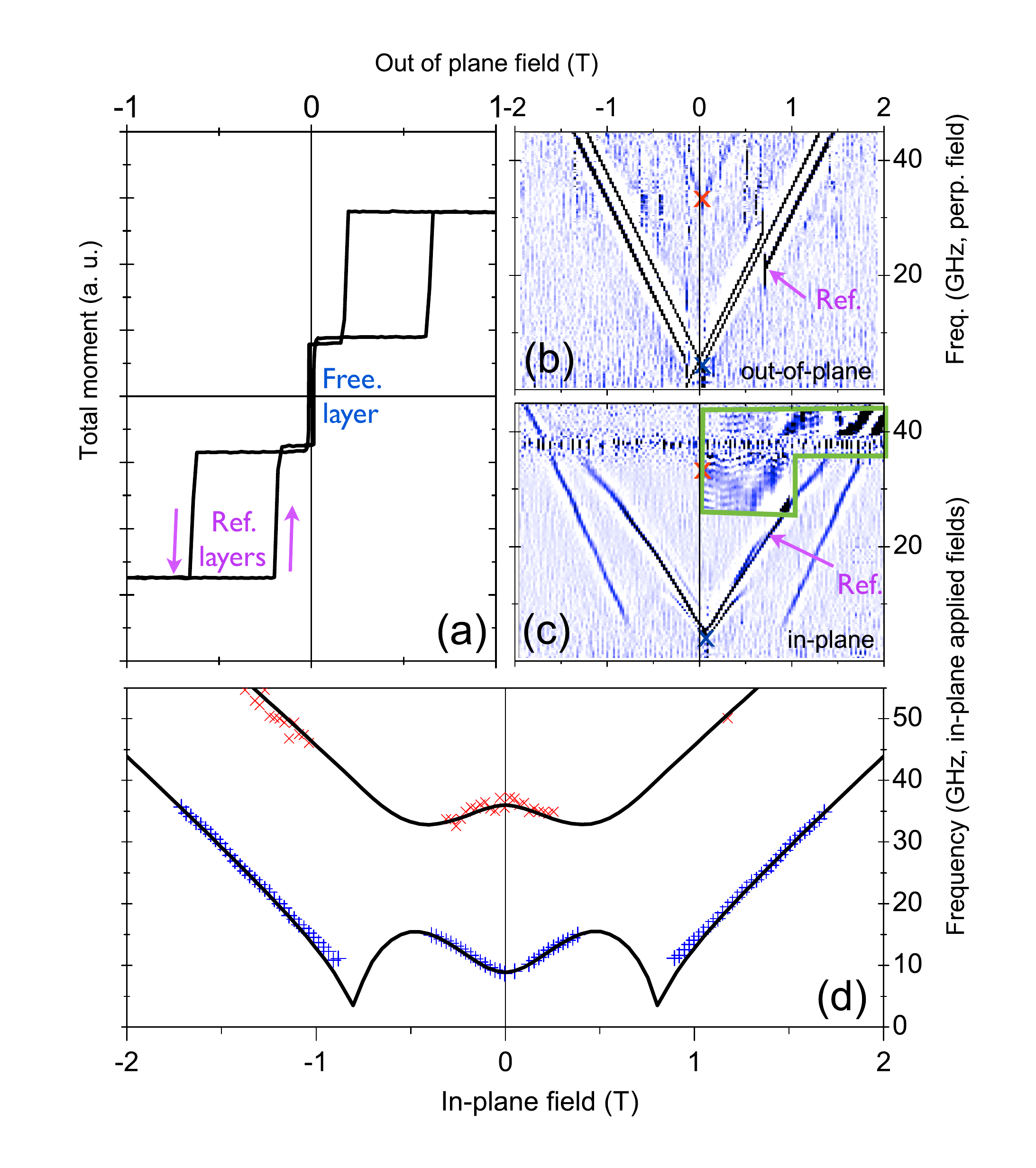}
\caption{(Color online). Properties of the MTJ with Mo 3\r{A} spacer. (a) Easy axis loop of the MTJ. (b) Permeability in the \{frequency-field\} planes for (b) out-of-plane field and (c) in-plane applied fields. The contrast has been strengthened in the green box. The violet arrows point at an eigenmode of the MTJ's reference layer. (d) Eigenmode frequencies along a hard axis loops (symbols) and modeling thereof (lines) with the parameters of Table 1.}
\label{figure_data}
\end{figure}

For Mo(5\r{A}), the magnetizations are found to be $M_{s1}=1.21\times10^6~\mathrm{A/m}$, $M_{s2}=0.763\times10^6~\mathrm{A/m}$ for the FeCoB layer and the [Co/Ni] multilayer, respectively. For the thinner spacers, the two parts of the free layer are sufficiently coupled to switch as a single block during easy axis loops, while the SAF reference system switches in two steps at much larger positive and negative fields [Fig.~\ref{figure_data}(a)]. Hard axis loops (not shown) comprise also the signals of the SAF of the MTJ such that the informations are too intertwined for a separate identification of the contribution of each magnetic sub-block.

To see to what extent the two parts of the free layer fulfill their roles, we measured the system's eigenexcitations, in addition its ground state determined by magnetometry. We illustrate our method in Fig.~\ref{figure_methods}. We are concerned by hybrid free layers with PMA and full remanence, i.e. in which the magnetizations of the two layers at remanence are collinear to the out-of-plane axis $(z)$. In this case the acoustical and optical eigenmode frequencies of the free layer under out of plane applied fields \cite{devolder_ferromagnetic_2016} are $f_{1,\,2}^\mathrm{exp} = \frac{\gamma_0}{2\pi} (H_{1,~2}^\mathrm{soft} \pm H_z)$ where the fields add for magnetizations up [i.e. along $+(z)$] and subtract for magnetizations down. The softening fields read \cite{devolder_ferromagnetic_2016}:
\begin{equation}
H_{1,~2}^\mathrm{soft} = \frac{{H_{k1}^{\mathrm{eff}}}+{H_{k2}^{\mathrm{eff}}}}{2}     + \frac {J } {2  {M_{s2}}  {t_2}}  + \frac {J } {2  {M_{s1}}  {t_1}}\pm \frac {\sqrt{\Delta}}    {2 {M_{s1}} {M_{s2}} {t_1} {t_2}}
\label{root}
\end{equation}

where $\Delta =2 J {M_{s1}} {M_{s2}} {t_1} {t_2}    ({H_{k2}^{\mathrm{eff}}}-{H_{k1}^{\mathrm{eff}}}) ({M_{s1}} {t_1}-{M_{s2}}    {t_2})+{M_{s1}}^2 {M_{s2}}^2 {t_1}^2 {t_2}^2    ({H_{k2}^{\mathrm{eff}}}-{H_{k1}^{\mathrm{eff}}})^2+J^2 ({M_{s1}} {t_1}+{M_{s2}}    {t_2})^2$. 
We have used the notation $H_{k1,~2}^{\mathrm{eff}}$ for the effective anisotropy fields of FeCoB (layer 1) and Co/Ni (layer 2), and $J$ for their interlayer exchange coupling.

%

When later conducting FMR versus $H_z$ experiments on the free layer (Fig.~\ref{figure_data}), we will confirm the finding of two V-shaped lines, with the apex of the V at $f_{1,\,2}^\mathrm{exp}$ and slopes $\pm \frac{\gamma_0}{2\pi} $. Unfortunately, the two data ($f_{1}^\mathrm{exp}$ and $f_{2}^\mathrm{exp}$) are insuffisient to determine the three unknown parameters $\{J,  {H_{k1}^{\mathrm{eff}}}, {H_{k2}^{\mathrm{eff}}}\}$. We have illustrated this problem in Fig.~\ref{figure_methods} (left panels) by choosing three different triplets $\{J,  {H_{k1}^{\mathrm{eff}}}, {H_{k2}^{\mathrm{eff}}}\}$ that yield the same calculated eigenexcitation frequencies in the $H_z$ field configuration (see the red and blue lines common to all left panels). 

\begin{table*}[!th]
\caption{Properties of the free layer subsystems. The magnetizations were fixed at $M_{s1}=1.21\times10^6~\mathrm{A/m}$, $M_{s2}=0.763\times10^6~\mathrm{A/m}$. The $\dagger$ symbol emphasizes the number that come from high field extrapolation when the dispersion curve departs from linear behavior at low fields. The linewidths are calculated from out-of-plane measurements of the lowest and highest frequency eigenmodes. In the absence of coupling, they would be indicative of the damping of the FeCoB  and Co/Ni systems, respectively. 
}\label{table}
\begin{tabular}{| c | c | c | c | c | c | c | c | c | c |}
  \hline
  & $f_1^\mathrm{exp}$  & $f_2^\mathrm{exp}$ & Fe$_{60}$Co$_{20}$B$_{20}$  & [Co3\r{A}/Ni6\r{A}]$_{\times 4}$ & Coupling  & $\frac{1}{2} \frac{\partial \Delta f }{\partial f}$ & $\frac{1}{2} \frac{\partial \Delta f }{\partial f}$ & TMR & RA \\ 
 & (GHz) & (GHz) & $\mu_0 H_{k1}^{\mathrm{eff}}$ & $\mu_0 H_{k2}^{\mathrm{eff}}$ & $J$ (mJ/m$^2$)  & FeCoB & Co/Ni &  &\\ 
Spacer  & & & $\pm$ 30 mT & $\pm$ 50 mT & $\pm$ 0.01  & $\pm 0.001$ &  &  & $\Omega.\mu \mathrm{m}^2$\\  \hline 
 Ta 3\r{A} & 1.0 & 24.5  & -20 & 850 &  0.07 &  0.009 & 0.03 $\pm 0.004$ & 137\%& 7.0\\ \hline 
  Mo 3\r{A}  & 8.6 & 35 & -50 &  920 &  0.58 & 0.010 & 0.03 $\pm 0.006$ & 137\%  & 8.7 \\ \hline 
W 3\r{A}   & -0.7 $\dagger$ & 27.6  & -190  & 890  & 0.22 &0.014 & 0.030 $\pm 0.003$ & 137\%  & 8.7 \\ \hline 
\end{tabular}
\end{table*}

To get an overdetermined problem, we performed in addition FMR vs in-plane field. Indeed as the two parts of the free layer have distinct anisotropies and thickness, their magnetizations tilt at different angles and the dynamics gets very sensitive to the exchange coupling $J$. The fitting procedure can be summarized this way: we first apply Eq. \ref{root} on the two remanent frequencies $f_{1,\,2}^\mathrm{exp}$ to get all the triplets $\{J,  {H_{k1}^{\mathrm{eff}}}, {H_{k2}^{\mathrm{eff}}}\}$ compatible with the easy axis field data. To determine which triplet is the correct one, we numerically calculate the magnetic configuration and eigenexcitations in in-plane applied field for all possible triplets, (Fig.~\ref{figure_methods}, right panels) and select the triplet that best match with the experiments. \\
In the absence of coupling [Fig.~\ref{figure_methods}, top middle panel], the hard-axis field eigenmode lines are W-shaped with two distinct softening fields at $H_x=\pm H_{k1}^{\mathrm{eff}}$ for the FeCoB layer and at $H_x=\pm H_{k2}^{\mathrm{eff}}$ for the Co/Ni multilayer. For finite coupling $J$, an anticrossing appears in the dispersion curves and there remains only one softening field per applied field sign. The layers still tilt at a different pace with the field. For large coupling (bottom panels, $J=0.9 ~\mathrm{mJ/m}^2$), the layers essentially tilt together and behave like a single unit. 

In practice the shape of the in-plane field FMR curves is a way to select the set of material parameters that best describe a sample. Besides, the linewidth of each mode in the out-of-plane field configuration can be used to estimate the damping of the two parts of the free layer. Indeed we will see that the coupling is small enough so that the lowest mode linewidth reflects the value of the FeCoB damping, while the linewidth of the highest frequency mode reflects the damping of the Co/Ni part of the free layer (for a detailed justification see for instance Figs.~6 and 7 in ref. \onlinecite{devolder_ferromagnetic_2016} and the related analysis).

The free layer eigenmodes were determined experimentally using Vector Network Analyzer FerroMagnetic Resonance (VNA-FMR~\cite{bilzer_vector_2007}). We applied fields either in the plane of the sample or perpendicularly to it.  
Thanks to the very different anisotropies of the magnetic subsystems within the MTJ, the eigenmodes have well separated frequencies [Fig.~\ref{figure_data}(b) and (c)]. There is one eigenmode undergoing sudden frequency jumps at \textit{exactly} the two switching fields of the SAF reference layers of the MTJ, but not undergoing any change at the free layer coercivity. The frequency of this mode is independent of the free layer inner spacer (i.e. Mo, W or Ta). This mode can thus be assigned to the other part of the MTJ (i.e. to the SAF reference system), and we shall thus not consider hereafter. Following the methods described earlier (Fig.~\ref{figure_methods}), a fit of the other modes to coupled macrospins was used to determine each layer's properties with conclusions gathered in Table I.


$f_1^\mathrm{exp}$ describes the strength of the effective fields that hold the softest part of the free layer. It is thus a relevant indication of its non volatility. The hybrid free layer with Mo spacer clearly outperforms substantially the other spacers in term of non volatility. Mo also ensures a low damping of the FeCoB layer at no expense of the overall anisotropy and of the transport properties. 
The main influence of the chemical nature of the spacer is the coupling strength (Table 1). The magneto-transport properties are insensitive to the spacer layer.
The couplings through the Ta spacer is low, and our experience is that it is weaker in our present Ni-terminated multilayers than when Co-terminated \cite{devolder_evolution_2016, le_goff_optimization_2015}. Comparatively to Ta, W 
could be considered as better owing to its larger interlayer exchange coupling. However, the W spacer has a clear detrimental effect on both the damping and the anisotropy of the FeCoB layer, which is consistent with other studies \cite{soucaille_probing_2016}. 


In summary we have studied hybrid perpendicular anisotropy free layers that couple a soft FeCoB layer with a  very anisotropic [Co/Ni] multilayer through various spacers made of refractory metals. The formerly used Ta(3\r{A}) spacer is outperformed by W(3\r{A}) and even more by Mo(3\r{A}) spacers, which combine the strongest interlayer exchange coupling (0.58 mJ/m$^2$) without sacrificing the anisotropies, the damping factors and the magneto-transport properties within the stack.

%

\end{document}